\documentclass[reprint,twocolumn,superscriptaddress,footinbib,aps,pre]{revtex4-1}

\usepackage[utf8]{inputenc}
\usepackage[english]{babel}

\usepackage{amsmath}
\usepackage{amsfonts}
\usepackage{amssymb}

\usepackage{tabularx}
\usepackage{booktabs} 
\usepackage{array} 

\usepackage[dvipdfmx, colorlinks, breaklinks,hyperfootnotes = false,citecolor = red]{hyperref} 
\usepackage[usenames]{color}

\usepackage{graphicx}

\begin{document}

\title{Encoding dynamics for multiscale community detection:\\ Markov time sweeping for the map equation}
\date{\today}
\author{Michael T. Schaub}\email[E-mail: ]{michael.schaub09@imperial.ac.uk}
\affiliation{Department of Mathematics, Imperial College London, London SW7 2AZ, United Kingdom}
\affiliation{Department of Chemistry, Imperial College London, London SW7 2AZ, United Kingdom}

\author{Renaud Lambiotte}
\affiliation{Department of Mathematics and Naxys, University of Namur, 5000 Namur, Belgium}

\author{Mauricio Barahona}\email[E-mail: ]{m.barahona@imperial.ac.uk}
\affiliation{Department of Mathematics, Imperial College London, London SW7 2AZ, United Kingdom}

\begin{abstract}
The detection of community structure in networks is intimately related to finding a concise description of the network in terms of its modules.  This notion has been recently exploited by the map equation formalism (M.~Rosvall and C.~T.~Bergstrom, {\em PNAS}, \textbf{105}(4), pp.~1118--1123, 2008) through an information-theoretic description of the process of coding inter- and intra-community transitions of a random walker in the network at stationarity.  However, a thorough study of the relationship between the full Markov dynamics and the coding mechanism is still lacking. We show here that the original map coding scheme, which is both block-averaged and one-step, neglects the internal structure of the communities and introduces an upper scale, the `field-of-view' limit, in the communities it can detect.  As a consequence, map is well tuned to detect clique-like communities but can lead to undesirable overpartitioning when communities are far from clique-like. We show that a signature of this behavior is a large compression gap: the map description length is far from its ideal limit. To address this issue, we propose a simple dynamic approach that introduces time explicitly into the map coding through the analysis of the weighted adjacency matrix of the time-dependent multistep transition matrix of the Markov process.  The resulting Markov time sweeping induces a dynamical zooming across scales that can reveal (potentially multiscale) community structure above the field-of-view limit, with the relevant partitions indicated by a small compression gap.
\end{abstract}
\maketitle

\section{Introduction}
The analysis of biological, technical and social networks has become extremely popular in recent years~\cite{Fortunato2010,Boccaletti2006,Arenas2008a}.
The availability of high dimensional relational data coupled with increasing computational power has set the ground for the investigation of complex systems from a network perspective, i.e., each agent or entity is viewed as a node interacting via multiple links with other nodes in the network.
Such a viewpoint aims to understand the global emergent behavior of the system from the interactions between the individual components of the system, in contrast to focusing on each part on its own.

In many cases of interest,  complex networks are far from being unstructured and contain relevant subgroupings or \emph{communities}, possibly organized into (not necessarily hierarchical) multiple levels~\cite{Simon1962}. The detection of such community structure can be of importance for the understanding of the interplay between the structural and functional features of the network. In particular, parts of the system operating on given scales could be represented with a simplified description at an appropriate level of coarse graining.

Community detection methods based on a variety of heuristics (including  modularity \cite{Newman2004,Newman2006} and spectral partitioning methods \cite{Fiedler1973,Fiedler1975,Shi2000,Kannan2000} among many others---see Refs.~\cite{Porter2009,Fortunato2010} for recent reviews) have been proposed to find an optimized split into communities. The communities thus found result from identifying groups with high intra-community weights as compared to the expected weights in surrogate models of the network.  In adopting such a structural criterion, these methods introduce an intrinsic scale that establishes limits on the communities they can detect, thus leading to potential misdetection \cite{Schaub2012}. Furthermore, such single scale methods are not suitable for the analysis of networks in which there is not a single `best' mesoscopic level of description, but rather multiple levels associated with different scales in the system~\cite{Simon1961}.

In order to account for the presence of multiple levels of organization, multiscale methods have been introduced that allow to search for the right scale at which the network should be analyzed \cite{Reichardt2004,Arenas2008,Ronhovde2009,Lancichinetti2011}. Recently, it has been shown that one can use the time evolution of a Markov process on the graph to reveal relevant communities at different scales in a process of dynamic zooming through the so-called \emph{partition stability}~\cite{Delvenne2010,Lambiotte2009,Schaub2012}. As the Markov time increases, the diffusive process involves multistep transitions and explores further afield the structure of the graph, resulting in the detection of community structure across scales, from finer to coarser. This dynamic approach has the advantage that it provides a unifying framework for structural community detection methods (such as modularity and spectral methods), which can be seen as particular cases of this approach involving one-step measures.

A different perspective is provided by an information theoretic framework that considers the problem of finding communities in a network as a coding or compression problem~\cite{Ziv2005, Rosvall2008, Rosvall2011, Raj2010}.
The underlying idea is that the presence of communities should imply the existence of an efficient and concise way to encode the behavior of a system in terms of its subgroups. Recently, the map equation method by Rosvall et al \cite{Rosvall2008, Rosvall2011} relies on a compression of the description length of a random walk inside and between communities to find good graph partitions. This method has received a lot of attention, since it has been shown to be extremely efficient on benchmark tests \cite{Lancichinetti2009} outperforming the popular modularity~\cite{Newman2004,Newman2006}. It has also been shown to be immune to the resolution limit~\cite{Fortunato2007} that affects the performance of modularity. However, the mathematical properties and possible limitations of the map equation remain relatively unexplored.

Here we show that the map equation can also be understood as a one-step method and, consequently, it suffers from an upper scale (the field-of-view limit) above which it cannot detect communities~\cite{Schaub2012}. This limited field-of-view can lead to overpartitioning when communities  are far from being clique-like \cite{Schaub2012}.  In addition, the one-step map coding scheme also neglects the internal structure of the communities and, in doing so, introduces a bias towards communities that are locally fast mixing (and in this sense clique-like).  We also show that the quality of the map partitioning can be assessed through the existence of a small compression gap, i.e., a small distance between the compression achieved by Map and its theoretical limit given by the true entropy rate of the Markov process. To alleviate some of these limitations, we introduce a dynamical approach that introduces time explicitly into the map coding scheme, by considering the time-dependent multi-step transition matrix of the Markov process on the network as the object of the map encoding. This introduces a dynamic zooming by sweeping through the Markov time,  which allows the detection of multiscale community structure with the map equation formalism.

\section{Community detection from a coding perspective: the map equation}

The map formalism considers the problem of partitioning a network into non-overlapping communities from a coding perspective. The original map formalism~\cite{Rosvall2008} equates the quality of the partition to the efficiency of a code that would describe the notional transitions of a random walker inside and between communities.  The Infomap algorithm can then be used to obtain good partitions through the optimization of this quality function. The underlying principle is that the code for such \textit{one-step} transitions of the random walker can be efficiently compressed in the presence of a strong community structure: short names for nodes (codewords) can be reused in different communities, much like street names can be reused in different cities of a country~\cite{Rosvall2008}.
In the original map equation, the movement of the walker is described in terms of two kinds of codebook. The first kind of codebook is specific to each community and assigns a unique codeword for each node inside it and a particular exit codeword for the community. An additional codebook contains unique codewords that describe the movements between different communities.
More recently,  a hierarchical extension of the map formalism (a recursive version of the original method) has been presented~\cite{Rosvall2011} as well as an extension for overlapping modules \cite{ViamontesEsquivel2011}.  We do not consider these extensions in detail here, as both methods are based on the same principles of the standard map equation and our findings are applicable to these as well.

\subsection{Definitions and notation}
An explicit rewriting of the original map formalism in terms of the stationary distribution of a random walk is as follows.  Consider a discrete time Markov process on a graph with $N$ nodes:
\begin{equation}
  \mathbf{p}_{k+1} = \mathbf{p}_k \, D^{-1}A \equiv \mathbf{p}_k \, M,
  \label{eq:randomwalk}
\end{equation}
where $\mathbf{p}_k$ is the $1 \times N$ (node) probability vector, $A$ is the (weighted) adjacency matrix of the graph, $D$ is the diagonal matrix containing the (weighted) degree of each node, and we have also defined $M$, the transition matrix of the random walk.  The stationary distribution of the random walk, $\mathbf{\pi}$, is then given by:
\begin{equation}
 \mathbf{\pi} = \mathbf{\pi} \, M.
 \label{eq:stationarypi}
\end{equation}

Consider now a partition of the network into $c$ communities indexed by $\alpha =1,\ldots,c$.
At stationarity, the probability of leaving community $\alpha$ (or of arriving at community $\alpha$) is
$$q_{\alpha\curvearrowright} = \sum_{i \in \alpha} \sum_{j \notin \alpha} \pi_i M_{ij}, $$
and  the overall probability of changing community is
$$q_{\curvearrowright} = \sum_{\alpha=1}^c q_{\alpha\curvearrowright}.$$
Similarly, the probability to stay within or to leave community $\alpha$  is $$p_\circlearrowright^\alpha = q_{\alpha\curvearrowright} + \sum_{i \in \alpha} \pi_i.$$

The map equation then defines the per-step description length of a code associated with this partition as:
 \begin{equation}
 L_M =  \sum_{\alpha=1}^c p_\circlearrowright^\alpha \, H(\mathcal P^\alpha) + q_\curvearrowright \, H(\mathcal Q),
 \label{eq:map}
\end{equation}
a weighted combination of the Shannon entropies:
\begin{align*}
H(\mathcal P^\alpha) &= -\dfrac{q_{\alpha\curvearrowright}}{p_\circlearrowright^\alpha} \log_2 \left( \dfrac{q_{\alpha\curvearrowright}}{p_\circlearrowright^\alpha}\right)
- \sum_{i \in \alpha} \dfrac{\pi_i}{p_\circlearrowright^\alpha} \log_2 \left( \dfrac{\pi_i}{p_\circlearrowright^\alpha}\right) \\
 H(\mathcal Q)  &= -\sum_{\alpha=1}^c \dfrac{q_{\alpha\curvearrowright}}{q_{\curvearrowright}} \log_2\left(\dfrac{q_{\alpha\curvearrowright}}{q_{\curvearrowright}}\right).
\end{align*}
The two terms in Eq.~\eqref{eq:map} correspond to two classes of codebooks that encode one-step transitions at stationarity viewed through the prism of the given partition. The first term stems from the ``community-centric" codebooks  with probability distributions $ \mathcal P^\alpha$ (and associated entropy) of being at or leaving from each of the communities.  The second term corresponds to the ``inter-community" codebook with distribution $\mathcal Q$ (and associated entropy) of changing community.

In the original map formalism it is proposed that a low $L_M$ is a characteristic of good partitions and the Infomap algorithm is used to search computationally for partitions with low $L_M$.

\begin{figure}[bt!]
 \centering
\includegraphics{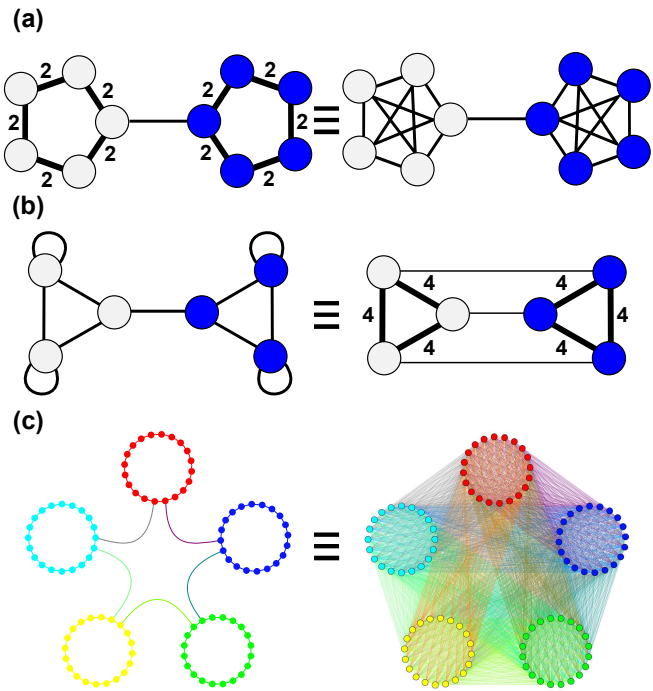}
 \caption{(Color online) \textbf{Equivalent graph partitions for the map equation.} Because the map equation ignores the specific connectivity of the graph, graph partitions with equal equilibrium and leaving probabilities become indistinguishable to Map. Different communities are represented by different colors (shades of gray). Unless indicated, the weight of the edge is 1. \textbf{(a)} Two graphs with different intra-community connectivity but the same map coding length, $L_M$. \textbf{(b)} Two graphs with different inter-community connectivity and the same $L_M$. \textbf{(c)} Two graphs with equal $L_M$ but very different inter- and intra-community connectivity. From the viewpoint of Map, a ring-of-rings is equivalent to a clique-of-cliques with different weights.}
 \label{fig:1}
\end{figure}

\section{Map encodes block-averaged, one-step transitions: implications for community detection}

As shown by the definitions above, the original map equation does not fully code for the dynamics of the Markov process~\eqref{eq:randomwalk}, as it only uses quantities derived from block-averaging of one-step transitions at stationarity. The simplifications involved in block-averaging the structure and in ignoring longer-term dynamics both have inter-related implications for community detection, which we now study in detail.

\subsection{Block-averaging the connectivity: the compression gap and a bias towards over-fitting to clique-like communities}

An examination of the terms in the map equation~\eqref{eq:map} reveals that the implicit block-averaging neglects the internal structure of the communities as well as the detailed inter-community connectivity. More precisely, given a particular partition, all graphs with the same equilibrium distribution $\pi$ and overall leaving probabilities $q_{\alpha\curvearrowright}$ will be indistinguishable in terms of their map quality, $L_M$, as exemplified in Figure \ref{fig:1}.

From the viewpoint of entropies, the map equation~\eqref{eq:map} is formally equivalent to a weighted sum of the entropies of i.i.d. stochastic processes with states visited according to: normalized ``community-centric'' probabilities
$\{ \{\pi_i/p_\circlearrowright^\alpha\}_{i \in \alpha}, q_{\alpha\curvearrowright}/ p_\circlearrowright^\alpha\}_{\alpha=1}^c$;
and normalized "leaving'' probabilities
$\{q_{\alpha\curvearrowright}/ q_{\curvearrowright}\}_{\alpha=1}^c$, respectively.
Alternatively, this procedure may be seen as formally equivalent to using a block-averaged transition matrix corresponding to a block-structured weighted (and in general directed) \textit{complete graph} with self-loops.  Consequently, Infomap exhibits a bias towards identifying communities that are formally equivalent to clique-like subgraphs.

In this sense, the map equation can be seen to code for a two-level, mean-field organization: one inside communities, one across communities. Such block-structured, all-to-all models are a good representation of community structure based on hierarchical cliques-of-cliques. Indeed, the map equation performs well in block-structured Erd\"os-Renyi benchmarks~\cite{Lancichinetti2009} and is not afflicted by the `resolution limit'~\cite{Fortunato2007}. On the other hand, there are important networks with a more marked local structure in which communities are not clique-like~\cite{Schaub2012}.  Because the map formalism has not been designed to detect such non clique-like communities with large effective distances, Infomap will tend to overpartition such networks.

\subsubsection*{Ignoring the detailed connectivity: the compression gap of the map equation}

The fact that Map ignores the detailed connectivity inside and outside the  communities leads to a sub-optimal coding scheme. This sub-optimality can be quantified through the \textit{compression gap} (defined below), which can be used as a measure of when the Map block-averaging assumptions are a valid simplification for the network under study.

Consider the Markov chain with transition matrix $M$ and stationary distribution $\pi$, as given in Eq.~\eqref{eq:randomwalk}.
The most efficient coding of the dynamics of the associated Markov process at stationarity is bounded from below by the entropy rate $h$~\cite{Shannon1948,Cover2006}:
\begin{equation}
\label{eq:h_def}
 h(\pi; M) = -\sum_{ij}\pi_i M_{ij}\log_2 (M_{ij}).
\end{equation}
The corresponding optimal encoding can be asymptotically achieved by endowing each node with a dictionary for its outgoing links, as shown by Shannon~\cite{Shannon1948}. This is a kind of `edge encoding.'

On the other hand, if we consider a coding scheme that gives each node a unique name within the whole graph, (i.e., a `node encoding'), then the corresponding coding length is bounded by the entropy rate of the i.i.d. random variable with probability distribution $\pi$, which is equal to the entropy of the stationary distribution:
\begin{equation}
H(\pi) =  -\sum_i \pi_i \log_2\left( \pi_i\right).
\end{equation}

The map coding scheme can be seen as a mixture of both:  it encodes nodes uniquely within communities, but encodes for transitions (`edges') between communities. Therefore, in general,
\begin{equation}
\label{eq:map_gap}
h(\pi; M) < L_M \leq H(\pi),
\end{equation}
and Map is sub-optimal in terms of its coding length~\footnote{Note that we do not have to construct a code to evaluate the map equation: one can find the optimal code-lengths from the entropy expressions or the map equation (\ref{eq:map}).}, as recognized by Rosvall and Bergstrom in their original publication~\cite{Rosvall2008}. This sub-optimality can be understood with a simple example: consider a community $\alpha_\text{out}$ from which there is only one possible link to another community $\alpha_\text{to}$. Map encodes this transition with two codewords: an exit codeword to signal the leaving of $\alpha_\text{out}$ and a codeword to identify the destination community $\alpha_\text{to}$. Clearly, the second codeword is redundant.

\begin{figure}[bt!]
 \centering
\includegraphics{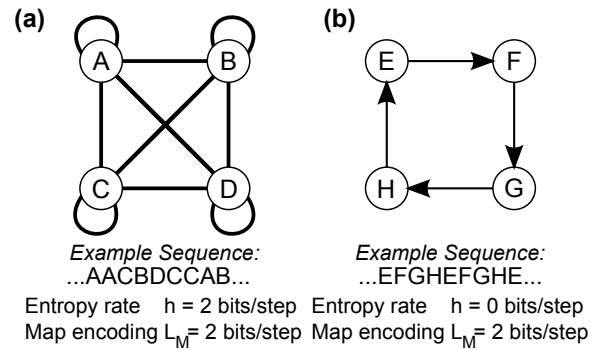}
\caption{\textbf{The compression gap of the map coding scheme.} \textbf{(a)}  For a clique with self loops,  the map coding scheme is optimal (assuming no communities) and is equivalent to a uniform i.i.d. process with four states. \textbf{(b)} In a directed cycle, the map coding is far from optimal. The movement of a random walker on this graph can be encoded by just denoting the starting position but the map coding scheme enforces unique names for each node and thus requires at least 2 bits/step. }
 \label{fig:2}
\end{figure}

Importantly,  if the graph is a weighted, directed clique (i.e., with transition matrix $M= \mathbf{1}\pi$), then
$h(\pi; M) = H(\pi)$ and the two coding schemes (`edge' and `node') give the same result (see Figure~\ref{fig:2}(a)). Therefore, the sub-optimality of the map coding is minimal when the graph is close to a clique. Consequently, the minimization of the map cost function is well suited to identify community structure that is a clique of cliques:  within each community Map uses a `node' encoding while between communities Map encodes transitions by default. In such a scenario, the map coding scheme is nearly optimal and close to the entropy rate.

The sub-optimality of the map encoding plays a significant role when encoding communities with restricted connectivity. For instance, if the community is a ring, a random walker has only two possible nodes to transition to, instead of $n_\alpha$  as assumed by Map. In this case, there is a large gap between the map description length $L_M$ and the optimal limit established by $h(\pi;M)$, indicating that the full consideration of the graph structure in the Markov dynamics could be exploited for a better encoding (see Figure~\ref{fig:2}(b) for an example).

This discussion highlights the fact that the block-averaging implicit in the original map scheme leads to a sub-optimality of the proposed map coding scheme that becomes significant when the network cannot be well described as a clique-of-cliques. In order to quantify this effect, we define the \emph{compression gap}, $\delta$:
\begin{equation}\label{eq:compression_gap}
\delta = (L_M-h)/h,
\end{equation}
which measures how close the map encoding is to optimality.  Note that other measures for the compression gap, such as
$\delta' = (L_M-h)/(H-h)$,
could be used and may be more suitable, or sensitive, in some cases. In this manuscript we stick mostly to the slightly simpler expression of $\delta$, as it is sufficient for our purposes. The compression gap can be used to establish when the communities identified by Map are far from being clique-like and hence serves as an indicator of the reliability of the partitions obtained by Infomap, as shown below.

\subsection{One-step transitions: the field-of-view limit and a bias towards overpartioning of non clique-like communities}\label{sec:quantitative:analysis}
As discussed above, the original map formalism is based on an implicit clique-like concept of community, and a community structure as a (statistical) clique of cliques. Although this model has proved successful in a variety of fields~\cite{Lancichinetti2009,Karrer2011}, relevant technological, social and biological networks are far from being clique-like~\cite{Schaub2012}.  In such cases, Infomap might tend to overpartition communities as a result of an upper scale (the `field-of-view limit') which stems from map encoding only for one-step transitions at stationarity. This field-of-view limit affects all one-step methods, including not only Map but also modularity. The field-of-view occurs on the opposite end of the well-known resolution limit that appears as a lower scale for modularity~\cite{Fortunato2007} but does not seem to impact Map~\cite{Lancichinetti2011}.

\subsubsection*{Overpartitioning of lattice-like graphs}
\label{sec:overpartitioning}
The overpartitioning induced by the field-of-view limit can be understood analytically through the following simple examples of lattice-like graphs.

First, consider a cycle graph of length $N$ with unweighted edges. The equilibrium distribution of the random walk on this graph is $\pi_i = 1/N, i=1,\ldots,N$. This graph has no community structure and the only relevant partition should be the global ``all-in-one".

For a partition of the ring into $c \geq 2$ communities indexed by $\alpha$ we have:
\begin{equation*}
\left \{q_{\alpha\curvearrowright} = 1/N, \, \forall \alpha; \quad q_{\curvearrowright} =c/N;
\quad p_\circlearrowright^\alpha = (n_\alpha +1)/N \right \}
\end{equation*}
where $n_\alpha$ is the number of nodes in community $\alpha$ and, clearly,  $\sum_{\alpha = 1}^c n_\alpha = N$.
The map cost function of this partition is
\begin{equation}
\label{eq:map_cycle}
L_M (\{n_\alpha\}_{\alpha=1}^c) = \frac{c}{N} \, \log_2 (c) + \sum_{\alpha = 1}^c \frac{n_\alpha +1}{N} \log_2 (n_\alpha +1).
\end{equation}
Using convexity arguments, it is easy to show that for a given $N$ and $c$, the minimal $L_M$ is attained for the partition with equally-sized communities with $n_\alpha =N/c, \, \forall \alpha$, if it exists.
For such a partition, the map equation~\eqref{eq:map_cycle} becomes:
\begin{equation}
\label{eq:ring_cut}
L_M (\{N/c\}_{\alpha=1}^c)
= \left(1+\frac{c}{N}\right)\log_2(N+c) - \log_2(c),
\end{equation}
with $c\geq 2$.  The case $c=1$ is the trivial ``all-in-one partition'' with $L_M(\{N\}_{\alpha=1})= \log_2 N$.

The relevant Map optimization for the cycle graph of size $N$ is then equivalent to finding which of the equal partitions into $c$ communities has the lowest $L_M$: $$ \underset{c}{\min} \, L_M (\{N/c\}_{\alpha=1}^c).$$
Assume $N/c$ to be real to facilitate the analysis, a relaxation which our numerics show not to affect the result. Then the partition with minimal $L_M$ has equal communities of size $N/c^*$ with $c^*(N)$ given by:
\begin{equation}
\label{eq:min_cycle}
\ln (N+c^*) =\frac{N}{c^*}-1 \quad \quad c^* \geq 2.
\end{equation}
It is easy to show that, for a long enough ring, such a partition will have lower $L_M$ than the `all-in-one' partition.  Indeed, the map equation partitions all cycles with $N \geq 10$.

Similar results are obtained for the regular $k$-cycles used as the starting point for the small-world construction (see Section \ref{sec:SW}). In this case, Map partitions the $k$-cycle into equally-sized communities of size $N/c^*$ given by:
\begin{equation}
\label{eq:min_pristine}
\ln \left(\frac{2N}{k+1}+c^*\right) =\frac{2N}{c^*(k+1)}-1 \quad \quad c^* \geq 2.
\end{equation}

The same reasoning can be applied to a torus network, i.e., the cartesian product of two cycles of lengths $R$ and $r$, with $N=rR$. This graph can be thought of as the discretization of a 2-dimensional lattice with periodic boundary conditions. It is easy to show that the optimal \emph{radially symmetric} partition of the graph (with $R>r$) is into communities of size $N/c^*$ with $c^*$ given by:
\begin{equation}
 \ln(2R+c^*) = \dfrac{2R}{c^*} -1
\end{equation}
Therefore, as the size of the lattice $N$ increases, Infomap will partition the torus into smaller sections. Our numerical exploration shows that the above solution is a conservative estimate and the overpartitioning induced by Map is even more acute for the torus: As $N$ grows, other even smaller patch-like partitions are obtained by the Infomap optimization.

\section{A dynamical enhancement of the map scheme: Markov time sweeping for the map equation}

As discussed above, the original map equation does not fully account for the dynamics of the Markov process~\eqref{eq:randomwalk}, as it only uses quantities derived from block-averaged one-step transitions. Such a simplification is reasonable for clique-like communities, which exhibit a small compression gap and can be fully explored in one step. However, networks of interest sometimes possess a multi-scale, non clique-like community structure which will go unrecognized by the original map equation due to its intrinsic bias towards cliques and the ensuing field-of-view limit.

The limitations of the map equation in such scenarios can be overcome by adopting concepts from \emph{partition stability}, a recently introduced dynamical framework for community detection \cite{Delvenne2010,Lambiotte2009,Schaub2012}. The idea is to consider the time evolution of the Markov process as a means to unfolding systematically the graph structure at different scales. This \emph{Markov time sweeping}, which is equivalent to considering multi-step transitions, applies a natural zooming process (from small to large scales) to the network.  A key aspect of this approach is the systematic sweeping across scales provided by the dynamics, which minimizes the effects of the resolution and field-of-view limits. For an extended discussion, see \cite{Delvenne2010,Lambiotte2009,Schaub2012, Delmotte2011}.

This Markov time sweeping can be used to endow the map equation with a dynamic zooming that allows it to detect multi-scale community structure, with relevant partitions characterized by a low compression gap. For simplicity, consider the continuous version of the Markov process~\eqref{eq:randomwalk} associated with a graph with adjacency matrix $A$ on $N$ nodes:
\begin{equation}
 \dot{\mathbf{p}} = -\mathbf{p} \, D^{-1} L,
\end{equation}
where $\mathbf{p}$ is a $1 \times N$ vector of probabilities, $D$ is the diagonal matrix containing the weights of each node and $L=D-A$ is the graph Laplacian. It is easily verified that this continuous-time Markov process has the same stationary distribution as the discrete-time random walk~\eqref{eq:randomwalk}~\cite{Delvenne2010,Lambiotte2009}.

The analytical solution of this system leads us to consider the discrete-time process:
\begin{equation}
\label{eq:transition_matrix}
 \mathbf{p}_{k+1} = \mathbf{p}_k \, T(t),
\end{equation}
where $T_{ij}(t) = [e^{-t D^{-1}L}]_{ij}$ is the effective transition probability between nodes $i$ and $j$ after a (Markov) time $t$.
Within this framework, it is easy to see that the original map formulation considers the \emph{linearized} version of $T(t)$ evaluated at time $t=1$.
Consequently, the original map equation scheme is included as a particular case in our formulation and we can always recover the standard Map results under our scheme \footnote{More specifically, the standard map equation is exactly recovered at time $t=1$ when considering either a linearized version of the dynamics \eqref{eq:transition_matrix} or of the corresponding discrete-time dynamics \eqref{eq:randomwalk}. The results presented here consider the nonlinear transition matrix \eqref{eq:transition_matrix} and hence the results at time $t=1$ shown here can display small deviations from those obtained from the standard (i.e., linearized in our case) map scheme.}.

Our approach is then to use the map equation to analyze the community structure of the \emph{time-dependent weighted network} $D \, T(t)$ as a function of the (Markov) time, $t$.  As time grows, the transition matrix $T$ becomes less sparse and more clique-like, yet in a structured manner that reflects the community structure of the network~\cite{Lambiotte2009}.
Consequently, the leaving probabilities $q_{\beta\curvearrowright}(t) = \sum_{i \in \alpha} \sum_{j \notin \alpha} \pi_i T_{ij}(t)$
increase with increasing time; the cost for encoding distinct communities increases too; and
map tends to find coarser communities that can be better represented as cliques. More specifically:
\begin{itemize}
\item For $t \rightarrow 0$, the leaving probabilities go to zero and the map equation is minimized by setting each node in its own community, as can be easily verified.

\item For $t \rightarrow \infty$, we approach the limit of an i.i.d. random process, i.e., $T(t) \rightarrow \mathbf{1}\pi$, where $\mathbf{1}$ is the vector of ones. In this limit, the map encoding for the ``all-in-one'' partition is optimal, since it results in a description length which is equivalent to the entropy rate. More precisely, it is easy to see from Eq.~\eqref{eq:map_gap} that $\delta(t) \to 0$ as $t \to \infty$.

\item For intermediate times,  the Markov time acts as a natural resolution parameter and the partitions of the time-dependent weighted graph $D \,T(t)$ become increasingly coarser.  By following the time evolution, we can check whether a particular partition corresponds merely to a transient or whether it is persistent for a range of times.
\end{itemize}

Furthermore, the compression gap~(\ref{eq:compression_gap}) can be used as an information-theoretic indicator of the reliability of the partitions found by Infomap at different Markov times. As discussed above, a low $\delta$ is expected when the partition reflects a community structure close to that of a clique of cliques, thus conforming to the assumptions underlying the map formalism. Therefore,  low values of $\delta(t)$ can be used to indicate relevant map partitions and also to identify the existence of a multi-scale community structure in the network.

This Markov time sweeping brings to the map equation what the \emph{partition stability} offers to modularity \cite{Delvenne2010,Lambiotte2009}; namely, the possibility to use time as a means to scan naturally through the resolution of community detection (from fine to coarse) in a manner that is consistent with the Markov dynamics on the graph. From this dynamical viewpoint,  the standard map equation corresponds to a time-snapshot of the diffusion dynamics. Furthermore, this dynamical approach is a natural framework for the map scheme, since it introduces a time-dependent but finite probability of jumping from any node to any other node at all times, in line with the formalism underpinning the map equation.

\section{Some illustrative examples}
\label{sec:numerics}

In this section, we illustrate the use of \emph{Markov sweeping map} with simple examples.
The procedure is as follows: For each Markov time, we construct the time-dependent network defined by $D \,T(t)$. We then optimize the (time-dependent) map cost function using the implementation of Infomap for directed graphs found online at \url{http://www.tp.umu.se/~rosvall/}, slightly modified to enable self-loops in the graphs.
We only consider here undirected networks but the method can be extended easily to directed graphs when we allow for teleportation~\cite{Lambiotte2012}.
For all examples, 100 runs of the Infomap algorithm at each Markov time were used to find the optimal partition.

\subsection{A network without community structure: the cycle graph }
 \begin{figure}[htb!]
 \centering
 \includegraphics{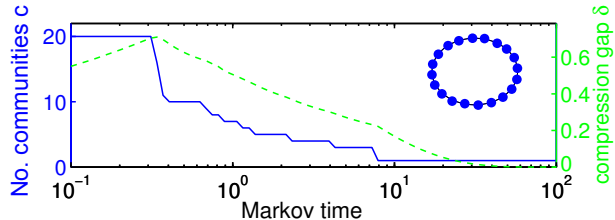}
 \caption{(Color online) \textbf{Markov sweeping map for a cycle graph with $N=20$.} As the Markov time increases, the map partitioning goes from the finest possible partition to the global `all-in-one' partition (solid blue line). However, as indicated by the featureless decay of the compression gap $\delta$ with no clear minima (dashed green line), no other relevant community is found between those two extreme partitions, thus signaling the lack of community structure.
In this case, optimizing the standard map equation finds 5 communities but a large compression gap $\delta \approx 2.48$ indicates that this partition is unreliable. Inset: analyzed graph.}
 \label{fig:3}
\end{figure}
As a first example, we apply Markov time sweeping to the ring network discussed in Section~\ref{sec:quantitative:analysis}.  Recall that for the cycle graph with $N=20$ nodes, our analytical arguments show that the original map scheme leads to a non-intuitive partition into 5 equal communities, instead of the expected `all-in-one' partition. However, the high compression gap of the 5-way partition found by the standard Map ($\delta \approx 2.48$) confirms that this partition is far from being formed by clique-like communities. Because standard Map is being applied to a network which does not conform to the implicit assumptions about community detection in the original map framework, we see an overpartitioning in this case.

As seen in Figure \ref{fig:3}, analyzing this network with the Markov-sweeping version of Map reveals  that there is no significant community structure in this graph.
Only the singleton partition (at very short times) and the global partition (at very long times) provide significant groupings of the nodes while all other partitions show high values of $\delta$.

\subsection{A simple network with multi-scale community structure}

\begin{figure}[htb!]
 \centering
 \includegraphics{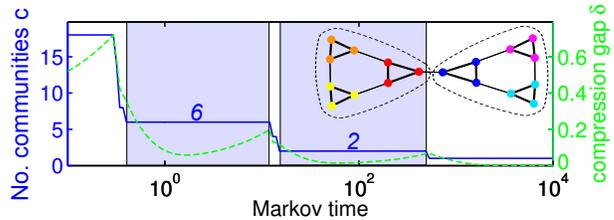}
 \caption{(Color online) \textbf{Markov sweeping map for a graph with a hierarchical community structure.}
 The graph analyzed (inset) has a clear community structure given by a hierarchy of triangles: the six smaller triangles (denoted by different colors) have edges within them of weight 100; they are grouped into two larger triangles with weaker links (the edges between the 6 small triangles have weight 10); the edge between the two big triangular structures has weight 1. The compression gap (dashed green line) shows two clear minima, indicating well defined partitions into 6 and 2 communities, corresponding to the two tiers of the hierarchy.  Standard Infomap finds only the 6 small triangles ($\delta \approx 0.62$). }
  \label{fig:4}
\end{figure}

Consider now a weighted graph with a distinct hierarchical community structure: two triangles of triangles with weighted links to reinforce the hierarchy (see inset of Figure \ref{fig:4}). In this example, the standard map equation method identifies the fine structure of six small triangles. (We note that the hierarchical map equation uncovers the two-tier hierarchy of communities in this graph.)

Our proposed Markov sweeping map also recovers the hierarchy of partitions across time-scales, as indicated by the sharp decreases in the compression gap $\delta$ when the six-fold and the two-fold partition are detected  (Figure \ref{fig:4}).
Our method also indicates over which timescales the relevant partitions appear to be natural. For instance, a change in the weights would induce changes in the lengths of the plateaux corresponding to the different levels of the hierarchy.  As stated above, hierarchical Map is able to resolve this clique-like community structure (while standard Map finds only the fine structure). However, if the multi-scale structure is not clique-like, hierarchical Map may fail to resolve the multi-scale structure, as shown in the network of small-world communities discussed in the next section.

\subsection{A ring of small-world communities}\label{sec:SW}
\begin{figure}[tb!]
 \centering
 \includegraphics{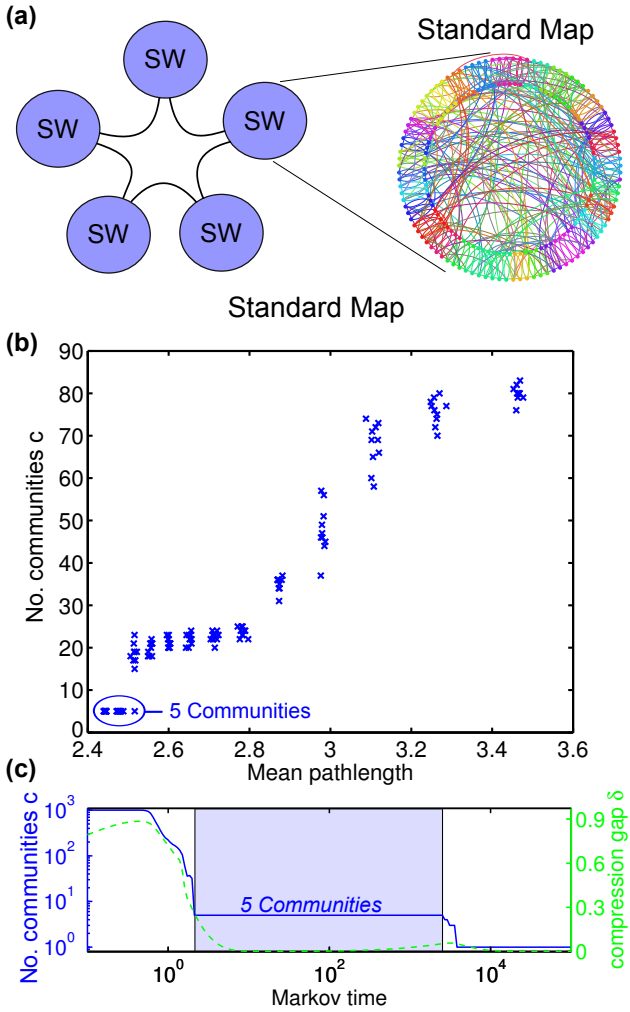}
 \caption{(Color online) \textbf{Community detection in a ring of small-world communities.} \textbf{(a)} Ring of 5 small-world communities with $N=200$ each. The edges within the small-worlds have weight 5 while the weight of the links between them is 1. All the small-worlds have an average number of randomly added shortcurts per node, $s$.   For $s=1$ (shown), standard Map shows a strong overpartitioning leading to an average of 16 communities inside each small-world (indicated by different colors in online version). \textbf{(b)} Number of communities found by standard Map \textit{vs.} mean pathlength inside the small-world communities.  The numerics shown correspond to 10 different realizations of the network with average number of shortcuts per node, $s=1,1.25,1.5,\ldots,3.75,4$. \textbf{(c)} Applying Markov sweeping map to the ring of small-worlds with $s=2.5$ (mean pathlength inside the small-worlds $\approx 2.7$) finds the relevant partition into 5 communities, while standard Map finds 23 communities in this case.}
 \label{fig:5}
\end{figure}

As a next scenario, we study a ring of five weakly connected small-world graphs~\cite{Watts1998} of 200 nodes each, as introduced in \cite{Schaub2012} (see Figure \ref{fig:5}(a)). We use the CONTEST toolbox \cite{Taylor2009} to generate small-world communities by adding random connections \`{a} la Newman-Watts \cite{Newman2000} starting from a pristine world with two nearest neighbours~\cite{Barahona2002} but
allowing for the possibility of multiple shortcuts at each node. 

As discussed in Section~\ref{sec:quantitative:analysis}, the standard map equation will tend to overpartition lattice-like structures, such as the pristine worlds ($k$-cycles) used as starting point for the small-world construction. As given by Eq.~\eqref{eq:min_pristine}, standard Infomap partitions the pristine world with $N=200$ and $k=2$ into 22 equally-sized communities.

This overpartitioning persists when few random shortcuts are added, as shown in Figure~\ref{fig:5}(b). Only when the average number of added shortcuts per node, here denoted by  $s$, is greater than 3.5 (and the mean distance within the small-world has become small) does standard Map obtain the right split into five communities. This is consistent with our discussion pertaining the field-of-view, i.e., the smaller the mean path length, the more clique-like the structure. In this case, hierarchical Infomap can even give a non-intuitive partition into 4 communities, due to the non clique-like nature of the communities. On the other hand, Figure \ref{fig:5}(c) shows that Markov sweeping allows Map to detect the relevant partition into 5 communities over an extended time-scale with a small compression gap.

\section{Discussion}
\begin{figure}[hbt!]
 \centering
 \includegraphics{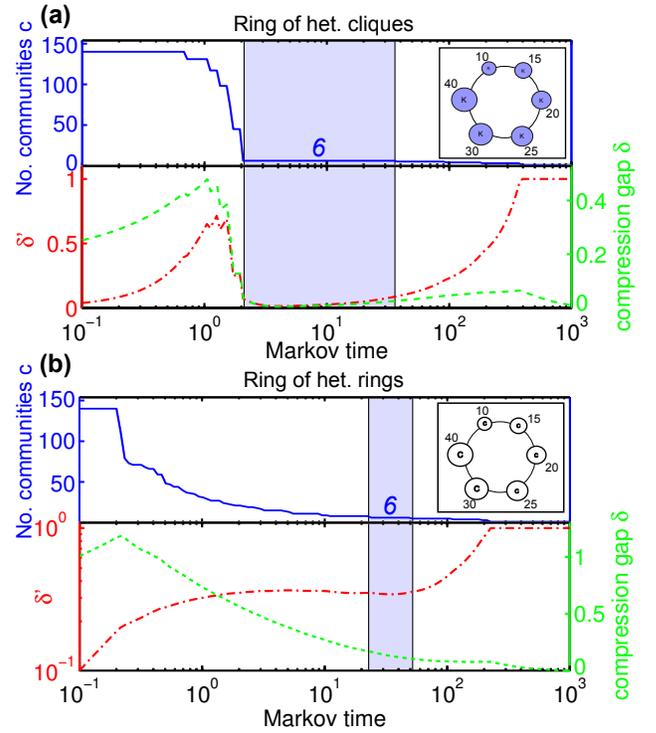}
 \caption{(Color online) \textbf{Community detection with heterogeneously sized subgraphs.} \textbf{(a)} Ring of 6 cliques with different sizes $\{10,15,20,25,30,40\}$. Upper panel: number of communities found by Markov sweeping map \textit{vs.} Markov time. Lower panel: both the compression gap $\delta$ (dashed green line) and the alternative compression gap measure $\delta '$ (red dashed-dotted line), clearly highlight the presence of a robust partition into 6 communities. Inset: analysed graph. \textbf{(b)} Ring of 6 rings with different sizes $\{10,15,20,25,30,40\}$. In this case the alternative measure $\delta'$ for the compression gap is better suited for the analysis, indicating the presence of the 6 rings by a relative minimum around Markov time 30. Inset: analysed graph. }
 \label{fig:6}
\end{figure}

A key insight to emerge from the map equation formalism is the fact that a coarse-grained description of a graph in terms of its communities is intimately related to finding concise descriptions of the information flow on these networks, and hence to the field of coding theory and data compression. However, the adoption of a coding or compression mechanism has important effects on the outcome of the algorithm and ultimately reflects the underlying assumptions about the concept of community.
Here we have shown that the original map equation formalism is inherently tuned towards a block-averaged notion of community structure as a weighted, statistical clique of cliques. This tuning stems from two inter-related simplifications: the block-averaged coding mechanism, which ignores the detailed connectivity and exhibits a large compression gap for non-clique structures, and the use of one-step quantities, which ignores the effect of multi-step flows in the communities and leads to an upper scale (field-of-view) for detection. This intrinsic bias of the map equation explains the excellent performance of the map equation in clique-like benchmarks but can lead to unexpected overpartitioning of networks if they differ strongly from the assumed clique-like organization.

We have shown that using the dynamical zooming provided by Markov time sweeping allows one to take into account multi-step flows and scan across all scales in a natural manner. The underlying idea is that, as time increases, the communities in the network will become more clique-like when analyzed through the  time-dependent weighted transition matrix of the Markov process. Therefore, the map formalism can be used to detect long-range communities as the Markov time increases, and the relevant communities will be signaled by a low compression gap. This Markov sweeping for the map equation can enhance the performance of the method by allowing it to detect non-clique communities and the presence of multi-scale community structure in networks. Importantly, the method still recovers all the results from the original map equation.

As stated above, the dynamic zooming across all scales effected by the Markov process is an integral ingredient of the method. Rather than just looking for the `right' scale, the community structure emerges from the integration of the information gathered systematically at all scales. This approach can help alleviate the reliance on a global scale which can affect the results when dealing with networks with communities with very heterogeneous sizes \cite{Lancichinetti2011a}.  In particular, Markov-sweeping map is able to detect heterogeneous cliques as obtained through the LFR benchmark, a fact consistent with the notion that cliques are all effectively one-step and that standard Map already performs effectively on such benchmarks (see also Figure \ref{fig:6} for an analysis with heterogeneously sized cliques).
Similarly our method performs well in detecting communities in a ring of rings with very dissimilar sizes as illustrated in Figure \ref{fig:6}, although when the heterogeneity of the relative ring sizes becomes very large, our approach will not identify all rings at once at the same level of the hierarchy. To improve further the applicability of the method to such problems,  one can use different dynamics for the Markov process \cite{Delvenne2012,Lambiotte2009}. This is an area of research we are currently pursuing.  However, since there is no community detection algorithm that will serve all purposes for all possible applications, one should complement the analysis with other methods based on different principles (e.g., local algorithms in those cases).

Adding a dynamical dimension to Map through Markov sweeping is just one of the possible ways to enhance the map equation and alternative approaches are worth pursuing. One direction would be the modification of the coding scheme. For instance, a more rigorous treatment would require to remove the constraint of having unique codewords within each community and allow also for encoding of walks instead of single step codewords. This generalization, however, would most likely lead to a breakdown of the simple coding picture that underpins the map equation.
Our work emphasizes the importance of the choice of dynamics on the network and shows that using a dynamical perspective may lead to a more natural framework for community detection, especially when the underlying system has an inherent flow.
In this paper, we have used the standard (unbiased) continuous-time random walk as a neutral first choice of dynamics. However, other continuous time or discrete time processes are possible (see also~\cite{Lambiotte2009} for a related discussion) in order to tune our community detection algorithm to different characteristics of the network.

Code is available online \footnote{\url{http://michaelschaub.github.com/MarkovZoomingMap/}}.
\vspace{0.5cm}
\begin{acknowledgments}\vspace{-0.5cm}
We thank J.-C. Delvenne for fruitful discussions, especially relevant for the quantitative analysis of the rings. We thank S.N. Yaliraki for helpful comments.
R.L. acknowledges funding from the Belgian Network DYSCO (Dynamical Systems, Control, and Optimization), funded by the Interuniversity Attraction Poles Programme, initiated by the Belgian State, Science Policy Office. M.B. acknowledges funding from grant EP/I017267/1 from the EPSRC (Engineering and Physical Sciences Research Council) of the UK under the Mathematics underpinning the Digital Economy program and from the US Office of Naval Research (ONR). The scientific responsibility rests with its authors.
\end{acknowledgments}

\bibliography{literature}

\end{document}